# Osmosis and thermodynamics explained by solute blocking

Peter Hugo Nelson
pete@circle4.com
Department of Physics, Benedictine University, Lisle, IL, USA

**Abstract**  A solute-blocking model is presented that provides a kinetic explanation of osmosis and ideal solution thermodynamics. It validates a *diffusive* model of osmosis that is distinct from the traditional convective *flow* model of osmosis. Osmotic equilibrium occurs when the fraction of water molecules in solution matches the fraction of pure water molecules that have enough energy to overcome the pressure difference. Solute-blocking also provides a kinetic explanation for why Raoult's law and the other colligative properties depend on the mole fraction (but not the size) of the solute particles, resulting in a novel kinetic explanation for the entropy of mixing and chemical potential of ideal solutions. Some of its novel predictions have been confirmed, others can be tested experimentally or by simulation.

## Introduction

Osmosis is a process that is fundamental to the physiology of all living things. It is the selective transport of water across a semipermeable membrane from high to low chemical potential caused by a difference in solute concentrations and/or hydrostatic pressures. This thermodynamic description is well established, but it says nothing about the kinetic mechanism responsible for osmosis. Despite its fundamental importance, the explanation for its physical basis has remained a controversial topic for well over a century, with many different mechanisms being proposed (Guell 1991; Weiss 1996). In current biophysics (Finkelstein 1987; Sperelakis 2012; Weiss 1996) and physics (Benedek and Villars 2000) textbooks, osmotic transport through a porous membrane is described as the *convective flow* of water through narrow pores that are selective for water over solutes. Within the convective flow model, a finite pressure gradient is always required within the pore for osmotic flow to occur.

Recently, a *diffusive* model of osmosis has been developed that is based on Fick's first law of diffusion (Nelson 2014, 2015). The diffusive model is conceptually distinct from the convective flow model (Kramer and Myers 2012; Kramer and Myers 2013; Sperelakis 2012). It is consistent with molecular dynamics simulations (Zhu et al. 2004b) of the motion of water molecules in aquaporins, which are integral membrane proteins that form water filled pores in the lipid bilayers of living things (Murata et al. 2000). The diffusive model is consistent with the observation that, in the absence of a water concentration difference, transport through the selectivity filter region of an aquaporin can be described as a continuous-time random walk (Berezhkovskii and Hummer 2002; Zhu et al. 2004b). As a result, permeation through aquaporins can be summarized by a knock-on jump mechanism (Hodgkin and Keynes 1955) and thus modeled using a framework wherein molecular transport is summarized by discrete jumps (Nelson 2012). The difference between the diffusive model and the convective flow model is exemplified by osmotic swelling/shrinking of a red blood cell within the constant-pressure Gibbs ensemble (Panagiotopoulos 1987). The convective flow model requires a finite pressure gradient within the pore, whereas the diffusive model does not.

The diffusive model of osmosis requires the use of an "effective water concentration" to be consistent with thermodynamics, but this concept was originally introduced without any kinetic justification (Nelson 2014, 2015). This paper presents a solute-blocking model of osmosis that overcomes that conceptual problem by providing a novel kinetic explanation of osmosis as a diffusive process that explains the origin of the "effective water concentration" concept. It successfully accounts for the single-file nature of osmotic transport though narrow pores and makes novel predictions that can be investigated experimentally and via computer simulation. Solute blocking also provides a simple kinetic explanation for the thermodynamics and colligative properties of ideal solutions. For dilute solutions the solute-blocking model simplifies to the diffusive model of osmosis, thus lending support to the simpler model.

## Solute Blocking Model

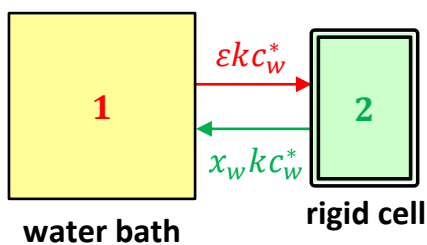

**FIGURE 1** FD diagram of the solute-blocking model of a rigid plant cell in contact with a bath of pure water. The water in the cell has a mole fraction $x_w$, which reduces the jump rate from box $2 \rightarrow 1$ by a factor $x_w$ compared with pure water. A positive hydrostatic pressure difference between the boxes reduces the jump rate from box $1 \rightarrow 2$ by an energy factor $\varepsilon$.

Figure 1 is a finite difference (FD) diagram (Nelson 2014, 2015) of osmotic permeation between a water bath and a rigid plant cell. The arrows indicate the unidirectional jump rates of water molecules between the two boxes within the Helmholtz ensemble (constant $T, V$) (Nelson 1998; Nelson et al. 1999). The primary purpose of this paper is to derive the form of the rate expressions in Fig. 1 using a novel solute-blocking model of osmosis and then to explore the predictions of that model. $k$ is the knock-on jump rate constant, $c_w^*$ is the concentration of pure water, $\varepsilon$ is the energy factor for jumps from box 1 → 2, and $x_w$ is the mole fraction of water in box 2. $\varepsilon = 1 - v_w \Delta p / k_B T$ is a linearized Boltzmann factor that accounts for a positive pressure difference $\Delta p = p_2 - p_1$ between boxes 1 and 2, where $v_w$ is the volume occupied by a single water molecule and $k_B T$ is the thermal energy.

The unidirectional jump rates associated with each arrow in Fig. 1 can be explained by the solute blocking illustrated in Fig. 2, which shows the reversible aquaporin state transitions that are possible when one end of the pore becomes temporarily blocked by a solute particle. Jumps from box 1 → 2 (states $a \to b$) are possible even if the pore entrance on the box 2 side is blocked. The reverse transition $b \to a$, a jump from box 2 → 1 is also possible, but once the selectivity filter is blocked on the box 2 side (state $a$), further jumps of water molecules from box 2 → 1 are not possible as indicated by the crossed out right-to-left blue arrow.

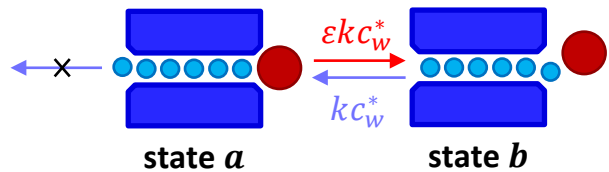

**FIGURE 2** Schematic diagram of two states of an aquaporin selectivity filter connecting the two boxes of Fig. 1. Random thermal motion produces a diffusive "knock-on" transition from state $a \to b$ wherein all the water molecules (small blue circles) move to the right as a water molecule enters the selectivity filter from box 1 and the far water molecule exits displacing the solution in box 2. The reverse transition from state $b \to a$ occurs when a water molecule enters the selectivity filter from box 2 and a water molecule is "knocked on" into box 1. However, once a solute (larger red circle) is located at the pore entrance (state $a$), further knock-on jumps into box 1 are blocked by the solute as indicated by the crossed-out arrow.

Figure 3 shows what happens if the boxes in the diffusive model are reduced to the size of a single water molecule with volume $v_w$. When the boxes are that size, they can be considered to contain either pure water, or a portion of a solute particle. As shown, box i contains pure water $c_i = c_w^*$ and box ii contains no water $c_{ii} = 0$ and jumps from box ii → i are blocked. The assumption is that there are only two possibilities for the system shown in Fig. 3: either there is a solute blocking the box ii pore entrance, or there is pure water next to it. In the first case (shown) a portion of a solute is occupying box ii and the aquaporin is blocked, but only for jumps in the ii → i direction. In the second case (not shown) box ii contains pure water and water permeation can proceed from box ii → i at rate $kc_w^*$. In order to determine the average unidirectional jump rate from box ii → i, we need the probability that a portion of a solute particle is occupying the water-sized box ii. If we make the ideal solution assumptions that the solute particles interact with the aquaporin entrance and with water molecules in a manner similar to water molecules, then the probability that a specific solute particle from macroscopic box 2 is filling nanoscopic box ii will be approximately proportional to the number of solute molecules. However, as shown in Fig. 3, only the left-most portion of the solute can block the pore entrance. As that blocking portion is approximately the size of a water molecule, the probability that a specific solute particle is filling nanoscopic box ii will be approximately the same as the probability that a specific water molecule is occupying nanoscopic box ii. If those are the only two choices, then the probability of a solute particle occupying box ii will be given by the solute mole fraction $x_s = 1 - x_w$, where $x_w$ is the mole fraction of water in macroscopic box 2. $x_w$ is also the probability that nanoscopic box ii contains pure water. Hence, a reasonable first-order approximation is that the size of the solute does not matter and the blocking probability is proportional to the mole fraction $x_s$ rather than the volume fraction of the solute in solution. As shown below, this distinction between mole fraction and volume fraction (or concentration) is central to the model correctly predicting the thermodynamics of osmosis and the other colligative properties.

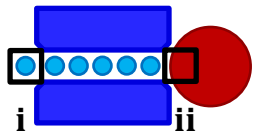

**FIGURE 3** Schematic diagram of an aquaporin selectivity filter separating two nanoscopic boxes (i and ii) that are the volume $v_w$ of a single water molecule. Box i is shown occupied by a water molecule and box ii contains a portion of a larger solute.

By inspecting Fig. 1, the condition for equilibrium is

$$\varepsilon = x_w. \tag{1}$$

Equation (1) provides a simple explanation of the origin of osmotic pressure and why it depends on the number – but not the size – of the solute particles. It indicates that two fractions are equal at equilibrium. $\varepsilon$ is the fraction of all the water molecules in box 1 that have enough energy to overcome the energy difference $\delta E = v_w \Delta p$, and $x_w$ is the fraction of all the particles in box 2 that are water molecules. Using the linearized energy factor, Eq. (1) yields the "Raoult's law version" of the van't Hoff equation

$$\Delta p = x_s c_w^* RT. \tag{2}$$

**Effective water concentration and diffusive model**

For dilute solutions, $n_s + n_w \approx n_w^*$, where $n_s$ and $n_w$ are respectively the number of moles of solutes and water molecules in volume $V$ and $n_w^*$ is the number of moles of pure water in the same volume $V$. Hence, using the definitions of mole fraction and concentration $x_s \approx n_s/n_w^* = c_s/c_w^*$, so that

$$x_s c_w^* \approx c_s, \tag{3}$$

where $c_s$ is the solute particle or osmotic concentration (osmolarity). Hence, the more accurate Eq. (2) reduces to the van't Hoff equation for dilute solutions, at equilibrium

$$\Delta p = c_s RT. \tag{4}$$

Substituting $x_s = 1 - x_w$ into Eq. (3) yields

$$x_w c_w^* \approx c_w^* - c_s = c_w. \tag{5}$$

Eq. (5) thus defines $c_w$ the "effective water concentration" in box 2 of Fig. 1 and provides a simple kinetic justification for its inclusion in the diffusive model of osmosis for dilute ideal solutions (Nelson 2014, 2015).

**Osmotic transport**

The model of Fig. 1 can be generalized by allowing box 1 to contain solutes and Eq. (5) can be used to define the effective water concentrations $c_{w_1}$ and $c_{w_2}$ in boxes 1 and 2, respectively. Solving the resulting model when $c_{w_1} \approx c_w^*$ and noting that $\varepsilon = 1 - \Delta p/c_w^* RT$ results in a molar flux of

$$j = -\mathcal{P}_f \left( \frac{\Delta p}{RT} - \Delta c_s \right), \tag{6}$$

where $\mathcal{P}_f = kV_2/A_2$ is the filtration permeability of the cell membrane, $k$ is the knock-on jump rate constant, $V_2$ is the volume and $A_2$ is the surface area of the cell and $\Delta c_s = c_{s_2} - c_{s_1}$ is the osmotic concentration difference between boxes 1 and 2. Equation (6) can be rewritten in terms of the volumetric permeation rate $Q$ yielding Starling's law of filtration

$$Q = -L_p(\Delta p - \Delta \pi), \tag{7}$$

where

$$L_p = \frac{\bar{V}_w A_2}{RT} \mathcal{P}_f = \frac{\bar{V}_w V_2}{RT} k \tag{8}$$

is the hydraulic permeability of box 2, $\bar{V}_w = 1/c_w^*$ is the partial molar volume of water, and the osmotic pressure difference $\Delta \pi = \pi_2 - \pi_1$ is defined as the equilibrium pressure difference. Each of the osmotic pressures is given by the van't Hoff equation

$$\pi = c_s RT. \tag{9}$$

Transient Eqs. (6) and (7) are identical to those of the traditional convective flow model and they successfully model experimental permeation through aquaporins (Mathai et al. 1996). When $\Delta\pi = 0$, Eq. (7) yields Darcy's law that fluid "flow" is proportional to the pressure gradient – despite the fact that the diffusive model is mechanistically distinct from the laminar flow assumed in the convective flow model of osmosis.

**Ideal solution thermodynamics**

Equilibrium equation (1) can be made more accurate by relaxing the assumption that the dimensionless energy step $\delta E/k_B T$ is small. In that case, the energy factor becomes a Boltzmann factor $\varepsilon = \exp(-\Delta E/k_B T)$ and Eq. (1) rearranges to

$$\bar{V}_w \Delta p + RT \ln x_w = 0, \tag{10}$$

which provides an alternate explanation of osmotic equilibrium in terms of two energies that cancel at equilibrium. The first term is the mechanical work $\Delta W$ done moving water from box $1 \to 2$ through a pressure difference $\Delta p$. The second term is the free energy decrease $-T\Delta S_{\text{mix}}$ when the water is "diluted" in box 2. Thus, the free energy change $\Delta F$ for water going from box $1 \to 2$ within the Helmholtz ensemble at equilibrium is

$$\Delta F = \Delta W - T\Delta S_{\text{mix}} = 0 \tag{11}$$

and the work done pressurizing the water is balanced by the entropy of mixing. Within the solute-blocking model, thermodynamic equation (11) is a direct consequence of kinetic equilibrium in the model system of Fig. 1.

When the system is not at equilibrium, Eq. (10) becomes

$$\mu_w = \mu_w^* + \bar{V}_w \Delta p + RT \ln x_w, \tag{12}$$

where $\mu_w$ and $\mu_w^*$ are the chemical potentials of the water in boxes 2 and 1 respectively. By comparing Eqs. (12) and (10), the chemical potentials are equal at equilibrium, i.e. $\mu_w = \mu_w^*$.

If box 2 is separated from box 1 and the pressure difference is relieved, then $\Delta p \to 0$ and

$$\mu_w = \mu_w^* + RT \ln x_w, \tag{13}$$

which is the chemical potential of water in an ideal solution.

**Raoult's law and colligative properties**

The solute-blocking model can also be used to explain Raoult's law and the other colligative properties of ideal solutions. The two-box system of Fig. 4 shows that for a gas-liquid system, dissociation jumps must occur from the liquid surface, and solute particles block water molecules from reaching a fraction $x_s$ of the surface from the liquid side. Modeling this situation with the solute-blocking model results in the FD diagram shown in Fig. 5. The evaporation rate is reduced from that for pure water by a factor of $x_w$ in analogy with the solute-blocking model of osmosis. The association rate for jumps from box $g \to w$ is not reduced by the presence of the solute because water molecules can condense on any portion of the liquid surface, including locations occupied by solutes.

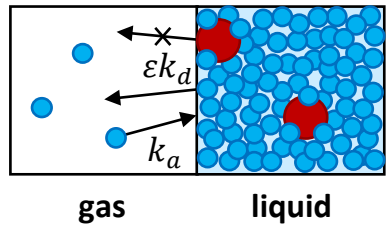

**FIGURE 4** Kinetic representation of the solute-blocking model of an aqueous solution in contact with its vapor (circles represent water molecules and solutes as in Fig. 2). Water molecules can only dissociate (evaporate) from the surface. Solute particles on the surface block evaporation as indicated by the crossed-out arrow. Water molecules in the gas can associate with (condense on) any portion of the liquid surface, including locations occupied by solutes.

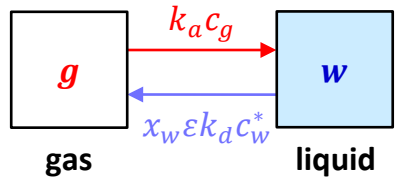

**FIGURE 5** FD diagram of the gas-liquid system of Fig. 4.

In the Gibbs ensemble (constant $T, p$) (Panagiotopoulos 1987), the energy factor for evaporation is

$$\varepsilon = \exp\left(\frac{-E_b - p\Delta v_{\text{vap}}}{k_B T}\right), \tag{14}$$

where $E_b$ is the binding energy of water molecules in solution and $p\Delta v_{\text{vap}}$ is the work done when a water molecule expands into the gas box at constant pressure. $\Delta v_{\text{vap}} = v_g - v_w$ is the volume change upon vaporization, which can be approximated by $\Delta v_{\text{vap}} \approx v_g$ as $v_g \gg v_w$ at normal temperatures and pressures. Hence,

$$\varepsilon = \exp\left(\frac{-E_b}{k_B T} - 1\right) \quad (15)$$

as for an ideal gas $pv_g = k_B T$. By inspecting the FD diagram, equilibrium occurs when

$$k_a c_g = x_w \varepsilon k_d c_w^*, \quad (16)$$

where $c_g = n_g/V_g = p/RT$. Substituting Eq. (15) into Eq. (16) results in Raoult's law

$$p = x_w p^*, \quad (17)$$

where

$$p^* = \frac{k_B T}{e v_0} \exp\left(\frac{-E_b}{k_B T}\right) \quad (18)$$

is the equilibrium vapor pressure of pure water at temperature $T$ and $v_0 = k_a v_w/k_d$. Eq. (18) can also be derived from the semi-classical partition function for a structureless ideal gas and an approximate partition function for an incompressible fluid (Baierlein 1999).

The remaining colligative properties can also be modeled using the solute-blocking model. Boiling point elevation results from the same mechanism as Raoult's law. As shown in Fig. 6, freezing point depression results because freezing is blocked by solutes whereas melting is not (Nelson 2014). The solute-blocking model thus provides a simple kinetic explanation for all of the colligative properties. The kinetics of these models can be investigated using molecular dynamics simulation techniques in a manner similar to kinetic models of ion channel permeation (Kasahara et al. 2013; Nelson 2011).

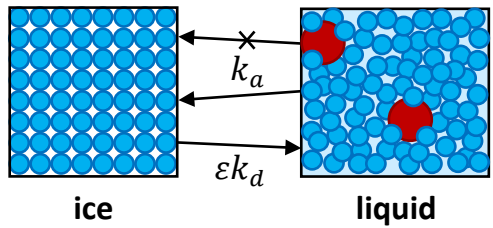

**FIGURE 6** Kinetic representation of the solute-blocking model of an aqueous solution in contact with pure ice (circles represent water molecules and solutes as in Fig. 2). In reality the two boxes are in direct physical contact, but they have been separated in the diagram to make room for the arrows indicating water molecules <u>a</u>ssociating with the ice (freezing) and <u>d</u>issociating (melting) at the interface. Water molecules can only associate (freeze) at the surface of the ice. Solute particles at the interface block freezing as indicated by the crossed-out arrow. Water molecules can dissociate (melt) from any portion of the ice surface, including locations covered by solutes.

**Tracer counterpermeation**

The solute-blocking model of osmosis also provides a simple kinetic explanation for the experimentally observed permeability ratio (Mathai et al. 1996). This can be understood by considering the counterpermeation eigenmode of tracer-labeled binary permeation (Nelson and Auerbach 1999a, b), in which there is only tracer-labeled water in box 1 with mole fraction $x_{w_1}$ and unlabeled water in box 2 with mole fraction $x_{w_2}$. With these boundary conditions, a kinetic analysis (Nelson 2002, 2014) of the state diagram of the $N_{sf} + 1$ occupancy states of an aquaporin selectivity filter with $N_{sf}$ single-file water molecules results in a unidirectional flux from box 1 → 2 that is given by

$$j_{1 \to 2} = \frac{\mathcal{P}_f \varepsilon x_{w_1} c_w^* \left(\frac{\varepsilon x_{w_1}}{x_{w_2}}\right)^{N_{sf}}}{\sum_{i=0}^{N_{sf}} \left(\frac{\varepsilon x_{w_1}}{x_{w_2}}\right)^i}. \tag{19}$$

The ratio of the unidirectional fluxes is given by

$$\frac{j_{1 \to 2}}{j_{2 \to 1}} = \left(\frac{\varepsilon x_{w_1}}{x_{w_2}}\right)^{(N_{sf}+1)}, \tag{20}$$

which corresponds to Hodgkin and Keynes' equation (10) with the concentration ratio replaced with the mole fraction ratio and the Boltzmann factor replaced with the energy factor (Hodgkin and Keynes 1955). If the osmolarity and pressure differences are zero, then $x_{w_1} = x_{w_2}$, $\varepsilon = 1$ and Eq. (19) reduces to

$$j_{1 \to 2} = \frac{\mathcal{P}_f x_{w_1} c_w^*}{N_{sf} + 1} = \mathcal{P}_d x_{w_1} c_w^*, \tag{21}$$

where $\mathcal{P}_d$ is the so-called "diffusive permeability" for tracer permeation. The resulting prediction for the permeability ratio, $\mathcal{P}_f/\mathcal{P}_d = N_{sf} + 1$, is consistent with experiment (Mathai et al. 1996) and molecular dynamics simulation (Zhu et al. 2004b). In contrast, continuum convective flow theory predicts that $\mathcal{P}_f/\mathcal{P}_d \sim 1$ for narrow pores (Finkelstein 1987).

## Discussion

The current biophysics (Finkelstein 1987; Sperelakis 2012) and physics (Benedek and Villars 2000) textbook model of osmosis is based on the assumption that permeation is always driven by a finite hydrostatic pressure gradient within the pore – even when there is no pressure difference between the boxes ($\Delta p = 0$). This internal pressure gradient is based on an argument that is equivalent to assuming that the pore contains bulk pure water and that each end of the pore is in local thermodynamic equilibrium with the exterior solution. As a result, there is always a pressure difference between the pore entrance and an adjacent solution having non-zero osmotic concentration. In contrast, the solute-blocking model is based upon the assumption that osmosis occurs by a diffusive process that can be summarized by knock-on jumps of water between the two boxes so that permeation is sorption-limited (Nelson and Auerbach 1999a, b) and no internal pressure gradient is required for osmotic swelling/shrinking of a red blood cell within the Gibbs ensemble.

The single-file nature of osmosis has been known since the late 1950s (Finkelstein 1987; Villegas et al. 1958) and a knock-on model of osmosis was proposed by Lea in 1963 (Lea 1963), but the *diffusive* knock-on model was rejected as a model of osmosis because the consensus view was that osmosis must be the hydrodynamic *flow* of water through a narrow pore driven by an internal pressure gradient (Finkelstein 1987). The present approach is based on the opposing view that osmosis is a *diffusive* process (Nelson 2015). Simulations of carbon nanotubes have already confirmed that a diffusive mechanism can explain water transport in the presence of a purely mechanical hydrostatic pressure difference $\Delta p$ for pure water (Zhu et al. 2004a), but simulations have yet to test the predictions of the solute-blocking model in the presence of an osmolarity difference.

Under tracer counterpermeation boundary conditions, the diffusive knock-on model is distinguished from continuum convective flow theory because it predicts a permeability ratio equal to $N_{sf} + 1$ that is consistent with the experimental value of 13.2 (Mathai et al. 1996). Molecular dynamics simulations (Zhu et al. 2004b) and the X-ray structure (Murata et al. 2000) of aquaporin-1 are consistent with $N_{sf} \approx 12$ single-file water molecules in the pore. The predictions of equations (19) and (20) have yet to be tested.

## Conclusion

The solute-blocking model provides a simple mechanistic explanation for why the mole fraction of water in a pressurized solution is equal to the Boltzmann factor for jumps from a pure water reference state at equilibrium. It also provides a kinetic explanation for the colligative properties of dilute solutions, the entropy of mixing, free energies, and the central role of the chemical potential in transport phenomena. It provides a new conceptual model that makes novel predictions for the microscopic details of osmosis and the other colligative properties that can be investigated experimentally and via computer simulation. A central theme of this paper is "thermodynamics from kinetics" (Nelson 2012). As we have seen, that philosophy can be used to address the long-running controversy surrounding osmosis.

# Acknowledgments

I wish to thank Eileen Clark, Robert Hilborn, Jaqueline Lynch, Philip Schreiner and the boys for helpful comments on an earlier draft of the manuscript. Support from the National Institutes of Health (Fellowship GM20584) and National Science Foundation (Grant No. 0836833) is gratefully acknowledged.